# A relationship between electron transfer rates and molecular conduction

Abraham Nitzan

School of Chemistry, The Sackler Faculty of Science, Tel Aviv University, Tel Aviv, 69978, Israel

## Abstract

This note discusses the relationship between a given intramolecular bridge assisted electron transfer rate and the corresponding zero bias molecular conduction of the same molecular species.

## 1. Introduction

Molecular electron transfer, one of the most basic chemical processes, has been an active field of research for over half a century.[1-6] Investigations of this reaction on the fundamental level focus on the rate of the transfer process between donor and acceptor species that exist in solution either as free solutes or as separate sites of a bigger molecule. In addition to rates, the yield of an electron transfer reaction is a meaningful observable when competing processes exist. Theoretical studies of these reactions aim to understand the interplay between solvent dynamical properties and between molecular structure and dynamics in determining these rates and yields.

Another manifestation of molecular electron transfer is gaining increasing attention in recent years: the passage of electronic current in a nano-junction comprised of two metal leads and a connecting molecule or a molecular layer, where the molecule has the role of a conducting component (See, e.g. Refs. 7-9). This leads to a new type of measurement where the observable is the molecular conduction (ideally of a single molecule), or more generally, the current-voltage characteristic of the molecular junction. On the theoretical side, these observables (and their dependence on the molecular structure and the molecule-metal bonding) are usually described using variants of the Landauer formula.[10-12] The latter is a formal relationship between the electron transmission coefficient associated with a given nano-structure (a property derived from scattering theory) and its conduction. The needed electron transmission properties of the molecular junction are computed following methodologies applied earlier in studies of 'conventional' molecular electron transfer,[13,14] using quantum chemical methods at different levels of approximation, or (for discussions of generic phenomenology) simples models, e.g., the super-exchange model[15] of the donor-bridge-acceptor (DBA) system.

Clearly, the conduction property of a given molecular system and the electron transfer properties of the same system should be closely related. One should keep in mind that because of tunneling there is always an Ohmic regime near zero bias. Obviously this conduction may be extremely low, indicating in practice an insulating behavior.[16] Of particular interest is to estimate the electron transfer rate in a given donor-bridge-acceptor (DBA) system that will translate into a measurable conduction of the same system when used as a molecular conductor between two metal leads. An earlier attempt in this direction[17] was limited to a 1-dimensional model and has disregarded the Franck-Condon factor in the electron transfer rate. In this paper I study this issue and derive an approximate practical relationship between these observables.

## 2. Theoretical considerations

Consider a DBA system, with a bridge that consists of N identical segments (denoted 1,2,...N) with nearest neighbor coupling $V_B$. For this case the non-adiabatic limit of electron transfer theory is usually valid, and the corresponding rate is given by

$$k_{D \to A} = \frac{2\pi}{\hbar} |V_{DA}|^2 \mathcal{F} \qquad (1)$$

where $V_{DA}$ is the coupling between the donor (D) and acceptor (A) electronic states and where

$$\mathcal{F} = \mathcal{F}(E_{AD}) = \sum_{\nu_D} \sum_{\nu_A} P_{th}(\varepsilon_D(\nu_D)) |\langle \nu_D | \nu_A \rangle|^2 \delta(\varepsilon_A(\nu_A) - \varepsilon_D(\nu_D) + E_{AD}) \qquad (2)$$

is the thermally averaged, Franck Condon (FC) weighted density of nuclear states. In Eq. (2) $\nu_D$ and $\nu_A$ denote donor and acceptor nuclear states, $P_{th}$ is the Boltzmann distribution over donor states, $\varepsilon_D(\nu_D)$ and $\varepsilon_A(\nu_A)$ are nuclear energies above the corresponding electronic origin and $E_{AD} = E_A - E_D$ is the electronic energy gap between the donor and acceptor states. In the classical limit $\mathcal{F}$ is given by

$$\mathcal{F}(E_{AD}) = \frac{e^{-(\lambda+E_{AD})^2/4\lambda k_B \Theta}}{\sqrt{4\pi\lambda k_B \Theta}} \quad (3)$$

where $k_B$ is the Boltzmann constant and $\Theta$ is the temperature, and where $\lambda$ is the reorganization energy. For the simple model where the bridge is described by a chain of N states, which are coupled to the donor and acceptor only via the first (1) and last (N) bridge levels, respectively, and when $E_D = E_A$, the non-adiabatic coupling takes the form

$$V_{DA} = V_{D1} V_{NA} G_{1N}(E_D) \quad (4)$$

so that

$$k_{D \to A} = \frac{2\pi}{\hbar} |V_{D1} V_{NA}|^2 |G_{1N}(E_D)|^2 \mathcal{F} \quad (5)$$

where $V_{D1}$ and $V_{NA}$ are the corresponding coupling matrix elements, and where $G_{1N}(E)$ is a matrix element of the bridge Green's function. In the tight binding approximation (nearest neighbor coupling) and in the weak coupling limit, $|V_B| \ll |E_B - E|$ it is given by

$$G_{1N}(E) = \frac{1}{E - E_N} \prod_{n=1}^{N-1} \frac{V_{n,n+1}}{E - E_n} = \frac{V_B^{N-1}}{(E_B - E)^N} \quad (6)$$

where $E_n$ and $V_{n,n+1}$ bridge energies and coupling elements, and where the second equality in (6) correspond to the situation where $E_n = E_B$ and $V_{n,n+1} = V_B$ for all bridge levels. The appearance of $\mathcal{F}$ in Eq. (1) indicates that the process is dominated by the change in the nuclear configuration between the two localized states of the electron.

Suppose now that the same DBA complex is used to connect between two metal contacts such that the 'donor' and 'acceptor' species are chemisorbed on the two metals, denoted 'left' (L) and 'right' (R) respectively. (In real molecular junctions these species correspond to the end groups of the molecular chain, but we continue to use the terms donor and acceptor in order to maintain the analogy to the corresponding electron transfer process). We wish to calculate the zero bias conduction $g$ of this junction and its relation to $k_{D \to A}$. First note that the conduction process does not involve localized states of the electron on the donor or the acceptor, so the factor $\mathcal{F}$ will not appear in $g$. We assume that at zero bias the metal Fermi energy $E_F$ lies in the gap between the HOMO and LUMO of the molecular bridge. In this case dephasing and energy losses arising from transient distortions of the bridge nuclear configuration are relatively small and will be disregarded. Assuming as before that states of the molecular complex are coupled to the metal only via the D and A orbitals, and that the latter are coupled only to their adjacent metal contacts, the conduction (at zero bias) can be derived from the weak coupling limit of the Landauer formula, or directly from the golden rule. It is given by[18] $g(E_F)$, with

$$g(E) = \frac{e^2}{\pi\hbar}|G_{DA}(E)|^2\,\Gamma_D^{(L)}(E)\Gamma_A^{(R)}(E) \tag{7}$$

Here $e$ is the electron charge and $\Gamma_D^{(L)}$ and $\Gamma_A^{(R)}$ are widths (imaginary parts of the corresponding self energies $\Sigma_D^{(L)}$ and $\Sigma_A^{(R)}$) of the $D$ and $A$ levels due to their couplings to the left and right metal leads, respectively.

Clearly, a relationship between $g$ of Eq. (7) and $k_{D\to A}$ of Eq. (5) can be established only if the electronic structure of the DBA species does not change considerably upon adsorption on the metal leads. In particular we assume that the weak coupling between the bridge and the rest of the (donor, acceptor and metals) system remains weak. In this case the Green's function element $G_{DA}$ is approximately related to the bridge Green's function $G_{1N}$ according to

$$G_{DA}(E) = \frac{\bar{V}_{D1}\bar{V}_{NA}}{\left(E - E_D - \Sigma_D^{(L)}(E)\right)\left(E - E_A - \Sigma_A^{(R)}(E)\right)}\bar{G}_{1N}(E) \tag{8}$$

The bars above $G_{1N}$ and the coupling elements correspond to the fact that their values for the chemisorbed molecule may be different from the corresponding values for the chemisorbed molecule. Eqs. (5), (7) and (8) thus lead to

$$\frac{g}{k_{D\to A}} = \frac{e^2}{2\pi^2}\frac{X_1 X_2 X_3}{\mathcal{F}} \tag{9}$$

where

$$X_1 = \frac{|\bar{V}_{D1}\bar{V}_{NA}|^2}{|V_{D1}V_{NA}|^2} \quad ; \quad X_2 = \frac{|\bar{G}_{1N}(E_F)|^2}{|G_{1N}(E_D)|^2} \tag{10}$$

and

$$X_3 = \frac{\Gamma_D^{(L)}\Gamma_A^{(R)}}{\left[\left(E_F - \tilde{E}_D\right)^2 + \left(\Gamma_D^{(L)}/2\right)^2\right]\left[\left(E_F - \tilde{E}_A\right)^2 + \left(\Gamma_A^{(R)}/2\right)^2\right]} \tag{11}$$

where $\tilde{E}_D$ and $\tilde{E}_A$ are respectively the donor and acceptor energies shifted by the real parts of the self energies Σ, and where all widths parameters Γ computed at $E_F$. Further simplification is obtained by introducing approximations. First, since the donor and acceptor species are chemisorbed on their corresponding metal contacts, their shifted energies $\tilde{E}_D$ and $\tilde{E}_A$ are expected to lie closer to the Fermi energies. We therefore assume that the denominator in (11) is dominated by the Γ parameters, i.e. $X_3 \approx 16/\left(\Gamma_D^{(L)}\Gamma_A^{(R)}\right)$. We also assume that this shift occurs uniformly in the DBA complex, without distorting its internal electronic structure. This implies that $X_1 \approx X_2 \approx 1$. These approximations may appear drastic, but they are sufficient for order of magnitude estimates as done below. For example, since $\Gamma_D^{(L)}$ and $\Gamma_A^{(R)}$ are of order ~1eV the error in the estimate of $X_3$ will be of order 1 and since far from resonance (which was assumed above) $G_{1N}(E)$ does not depend strongly on the energy the error in taking $X_1=1$ is similarly small. Eq. (9) then lead to

$$g \approx \frac{8e^2}{\pi^2 \Gamma_D^{(L)} \Gamma_A^{(R)} \mathcal{F}} k_{D \to A} \qquad (12)$$

## 3. Discussion and conclusion

The following observations can be made. First, the appearance of $\mathcal{F}$ in the denominator of Eq. (12) is a direct consequence of the fact that the reorganization energy, a principal factor controlling molecular electron transfer rates, does not affect the corresponding conduction. This results from the important physical difference between the two phenomena: the first involves, while the second does not involve, localization of the electron on the donor and acceptor species. Secondly, the appearance of the widths parameters $\Gamma_D^{(L)}$ and $\Gamma_A^{(R)}$ in the same denominator is less obvious and may appear counter-intuitive. It expresses the fact (see Eq. (8) and recall that $\text{Im}\Sigma = -\Gamma/2$) that the effective coupling between the bridge and the chemisorbed 'donor' and 'acceptor' levels decreases for increasing Γ. Finally, it may be shown by extending an argument given in Ref. 19 that, provided the energy spacing $E_B - E_F$ between the bridge levels and the Fermi energy is large relative to $k_B T$, Eq. (12) holds also when the electron transfer process involves thermal activation into the bridge states.

The parameters appearing in Eq. (12) may be estimated form experimental data or theoretical models. Using the classical expression for $\mathcal{F}$, Eq. (3), we have for $E_D = E_A$ $\mathcal{F} = \left(\sqrt{4\pi\lambda k_B T}\right)^{-1} \exp(-\lambda/4k_B T)$. For a typical value of the reorganization energy ~0.5eV, and at room temperature this is ~ 0.02(eV)$^{-1}$. Taking also $\Gamma_D^{(L)} = \Gamma_A^{(R)}$ ~0.5eV leads to $g \sim \left(e^2/\pi\hbar\right)\left(10^{-13} k_{D\to A}(s^{-1})\right) \cong \left[10^{-17} k_{D\to A}(s^{-1})\right]\Omega^{-1}$. This sets a criterion for observing Ohmic behavior for small voltage biases in molecular junctions. For example, with a current detector sensitive to pico-amperes, $k_{D\to A}$ has to exceed ~$10^6$s$^{-1}$ (for the estimates of $\mathcal{F}$ and $\Gamma$ given above) before measurable current can be observed at 0.1V voltage across such a junction. It should be kept in mind that this estimate was done using a value for $\mathcal{F}$ based on the high temperature approximation (3), and should modified in situations where $\mathcal{F}$ is dominated by high frequency ($\hbar\omega > k_B T$) intramolecular modes.

Eq. (12) provides a practical expression for making such estimates, but its use should be exercised with caution. This expression is a relation between the conduction of a molecular structure made of a bridge connecting two special chemical groups chemisorbed on corresponding metal leads and the electron transfer rate of the same molecular structure in solution, where the same special groups play the roles of donor and acceptor. Its derivation involves a drastic assumption that, apart from a uniform energy shift, the electronic structure of this molecular structure when connecting between the metal leads is not considerably different from the electronic structure of the corresponding species in solution. We note that most recent calculations of molecular conduction rely on similar assumptions, using as input electronic structural parameters of the free molecule (e.g., assuming that the molecular bridge is electrically neutral at zero bias). Another point to emphasize is that the result (12) corresponds only to conduction at zero bias. An important attribute of experimental setups involving molecular conductors is the possibility to employ their finite bias conduction, which is not directly related to the corresponding electron transfer property. Keeping these limitations in mind, Eq. (12) provides a useful approximate relationship between two extremely important molecular transport properties.

7**Acknowledgements.** This research has been supported by the U.S.-Israel binational Science Foundation. I thank D. Baratan and M. Ratner for helpful discussions.